\documentstyle[emulateapj,psfig,apjfonts]{article}

  \def\itm#1 {\vskip10pt \noindent \square\ {\bf #1} }
  \def\square {\hbox{\vrule width5pt height5pt}}

  \def\deg      {{\ifmmode^\circ\else$^\circ$\fi} } 
  \def\arcm    {{\ifmmode {'\ }\else$'     $\fi} } 
  \def\arcs    {{\ifmmode{''\ }\else$''    $\fi} } 



\newcommand{\mum}{$\,\mu$m}

\lefthead{Chapman et al.}
\righthead{HST images of submm galaxies}

\begin{document}
\title{{\it Hubble Space Telescope} images of 
submillimeter sources: large, irregular galaxies at high redshift}
\author{S.\,C.\ Chapman,$\!$\altaffilmark{1}
R.\ Windhorst,$\!$\altaffilmark{2}
S.\ Odewahn,$\!$\altaffilmark{2}
H.\ Yan,$\!$\altaffilmark{2}
C.\ Conselice,$\!$\altaffilmark{1}
}

\altaffiltext{1}{California Institute of Technology, MS 320-47, Pasadena, CA, 91125}
\altaffiltext{2}{Arizona State University, Dept.\ of Physics and Astronomy,
Tempe, AZ, 85287-1504}

\slugcomment{To appear in the Astrophysical Journal}


\begin{abstract}
We present new {\it Hubble Space Telescope} STIS, high-resolution
optical imaging of a sample of 13 submillimeter (submm) luminous
galaxies, for which the optical emission has been pinpointed either through
radio-1.4\,GHz or millimeter interferometry. We find a predominance of
irregular and complex morphologies in the sample, suggesting that mergers
are likely common for submm galaxies. 
The component separation in these objects are on average a factor
two larger than local galaxies with similarly high bolometric luminosities.
The sizes and star formation rates of the submm galaxies are consistent with
the {\it maximal} star formation rate densities of
20\,M$_\odot$\,kpc$^{-2}$ in local starburst galaxies (Lehnert \& Heckman 1996).
We derive 
quantitative morphological information for the optical galaxies hosting
the submm emission; total and isophotal magnitudes,
Petrosian radius, effective radius, concentration, aspect ratio, surface
brightness, and asymmetry. 
We compare these morphological indices with those of other
galaxies lying within the same STIS images. 
Most strikingly, we find $\sim$70\% of the submm galaxies to be extraordinarily
large and elongated relative to the field population, regardless of optical
magnitude.  
Comparison of the submm galaxy morphologies with those of optically selected
galaxies at $z\sim2-3$ reveal the submm galaxies to be a morphologically 
distinct population, with generally larger sizes, higher concentrations
and more prevalent major-merger configurations.
\end{abstract}



\section{Introduction}
\label{secintro}

Sub-mm galaxies (Smail, Ivison \& Blain 1997) are candidates for the 
progenitors to massive spheroids in the local Universe
(e.g., Lilly et al. 1999). Their
luminosities ($>10^{12}$L$_\odot$) imply star formation rates (SFRs) large
enough to build $>>$M$^*$ galaxies in much less than 
a Gyr, while their volume density (Barger, Cowie \& Sanders 1999; 
Chapman et al.~2003a) is
comparable to that of giant ellipticals locally. However, it is not at all
clear that submm galaxies do in fact build giant ellipticals. The
timescales for the huge luminosities could be very short (e.g., 
Blain et al.\ 1999, Smail et al.\ 2003), 
consistent with galaxies with lower masses than the
giant ellipticals. Tacconi et al.\ (2002) have shown evidence that
local galaxies of comparable luminosities to submm galaxies (Ultra-Luminous
InfraRed Galaxies -- ULIRGs) cannot evolve into giant ellipticals.
The merger of two gas rich disks in a system of even modest mass could
lead to the generation of a brief period of great luminosity
(Mihos \& Hernquist 1996).
Sub-mm galaxy selection is susceptible to picking out the most luminous
bursts at any epoch. 
However, detection of CO
molecular gas in 5 submm galaxies (Frayer et al.\ 1998,1999; Neri et
al.\ 2003) and a claimed strong clustering from a spectroscopic
study of the submm population (Blain et al.\ 2004) are
suggestive of massive systems for at least a fraction of the submm galaxies.
Locally, the most luminous objects are the
ultra-luminous infrared galaxies (ULIRGs), which show evidence for being
driven by mergers (Mihos \& Bothun 1998).

If the submm sources are snapshots of galaxies in the process
of formation, their optical morphologies should reveal the 
mergers in process.
However an ongoing difficulty with both the
identification of the submm galaxies (large submm beam sizes and thus positional
uncertainties), and the intrinsic faintness of most of the sources at 
optical wavelengths, has impeded a detailed morphological study of the
population (Smail et al.\ 1998).
The identification hurdle was overcome for the majority of the submm population
by using the deepest radio surveys 
(rms noise as low as 4\,$\mu$Jy at 1.4\,GHz -- Richards 2000, 
Ivison et al.\ 2002, Fomalont et al.\ in preparation) 
to pinpoint the submm
emission through a high-redshift extrapolation of
the far-IR/radio correlation (e.g., Helou et al.\ 1985, Condon 1992).
The faint microJansky radio source population is 
dominated by star-forming galaxies and low-luminosity active galactic nuclei
(AGNs), distant analogs
of local luminous infrared galaxies, with suggested
star-formation rates of 10-1000 $M_{\odot}$/yr (Windhorst et al.~1995;
Richards et al.~1998).
The submm galaxies constitute a subset of the faint radio population,
with 65\% of submm galaxies showing radio detections to 30$\mu$Jy at 1.4\,GHz
(Barger, Cowie \& Richards 2000; 
Chapman et al.\ 2001a, 2003b; Ivison et al.\ 2002).


The optical morphologies of the submm galaxies have been difficult to study at 
high resolution, 
%
owing in part to the difficulty of obtaining deep HST images over wide
areas (S$_{850 \mu m}>5$\,mJy submm galaxies number 500\,deg$^{-2}$). 
The few submm galaxies with HST observations (Smail et al.\ 1998,
Ivison et al.\ 2001; Chapman et al.\ 2002a,b; Sato et al.\ 2002) 
show a range in morphologies, with 
many apparently multi-component systems suggestive of
the {\it early stage mergers}, seen in $\sim$25\% of the
local ULIRGs (Goldader et al.~2001, Surace \& Sanders 2000). 
In general, local luminous galaxies appear to
encompass a luminosity-dependent morphology relation, from less
disturbed (Ishida \& Sanders~2001) to
major merger morphologies (Kim et al.~1998a,b; Goldader et al.~2001) as
luminosity increases, with the most luminous
($>10^{13}$L$_\odot$)
often dominated by QSOs (Farrah et al.\ 2002).
However, without a statistical sample of submm galaxies
with {\it HST} resolution, it is difficult to assess how the
submm galaxies relate to local luminous galaxies detected with {\it IRAS}.
In this paper, we present a morphological analysis of
a sample of radio-identified submm galaxies
using new deep {\it HST-STIS} observations in the rest-frame UV. 
We discuss the sample selection and
observations (\S~2), analyze the submm galaxies relative to the
general field population and the optically selected $z\sim3$ galaxies
(\S~3), and finally discuss the
generally peculiar morphologies found for the submm galaxies (\S~4).

\section{Sample and Observations}

We assembled a sample of 13 submm galaxies 
lying in four distinct fields (the HDF,
SSA13, SSA22 and Westphal-14hr) for followup with {\it HST}.
The submm galaxies were detected with SCUBA on the JCMT, and
pinpointed within ground-based
optical images using radio, X-ray, or millimeter
interferometric identifications.
These submm sources were first presented in 
Barger et al.\ (1999), Barger, Cowie \& Richards (2000), 
Chapman et al.\ (2000, 2001a, 2001b, 2002a, 2002c).
Details of each source and its identification at multiple wavelengths are
presented in Appendix~1.
All submm galaxies are brighter than S$_{850 \mu m}>4$\,mJy, implying
far-infrared luminosities greater than 10$^{12.5}$\,L$_\odot$ if they lie
at $z>1$. 
Eleven out of 13 sources have associated radio emission ($>5\sigma \times$ rms),
with X-ray emission serving as secondary positional identifier. 

These galaxy candidates were chosen 
from the catalogs in each field to uniformly sample the
$I$-mag distribution of the total radio-submm galaxy population.
In Fig.~1, we plot the $I$-mag distributions of the parent sample and the HST
sample. 
The sources lie in the range $I=21$--$26$ mag.
We have presented measurements for $I$-band fluxes if the
significance is 3$\sigma$ or higher, otherwise we have represented the
non-identification of the submm/radio source by the 2$\sigma$ limit
of the optical image.
Optical magnitudes were measured using 
the SExtractor program (Bertin \& Arnouts 1996),
from our own imagery as well as imagery taken from the archives using
(Westphal-14, WHT,
$I(5\sigma)=25.2$ (point source limit); 
SSA22, CFHT-12k, $I(5\sigma)=26.4$ (point source limit); 
HDF, Subaru/SUPRIME -- Capak et al.\ 2003; 
SA13, KPNO/MOSAIC and Subaru/SUPRIME archival imagery $I(5\sigma)=26.1$).
Ground based $I$-mag are measured 
in 3\arcsec\ aperture centered on radio position.
No reference was made in the selection 
to the ground-based optical imagery beyond the magnitude
measurements,
and the sample should be unbiased with respect to morphology.

HST imaging for this sample was obtained through a Cycle\,10 program with the
{\it Space Telescope Imaging Spectrograph} (STIS) to study the
morphologies of the submm luminous galaxies. Between one and three orbits 
of integration time (tailored to the ground-based magnitudes), 
giving 2340--7280\,sec of {\it LOW SKY} observation,
were split between two exposures per orbit, using the $50CCD$-clear filter. The
pipeline-processed frames were calibrated and aligned, and cosmic ray rejected,
using standard IRAF/STSDAS routines. 
The pixel size in the STIS images is 0.0508\arcsec. 
The $50CCD$-clear filter is roughly a Gaussian with 1840\,Angstrom half width
and a pivot wavelength of 5733\,Angstrom; we refer to the associated AB
magnitude as $R'(573)$. The sensitivity
limit reached is $R'(573)\sim27.1-27.6$ mag (5$\sigma$), 
corresponding to $R\sim27.6$--$28.1$ mag for a point
source with an Sb galaxy SED.
The STIS images are presented in Fig.~2, with radio centroids marked with a 
cross. All sources except one (SMM\,J131235.2) were significantly detected 
by STIS.

The astrometry in the small (50\arcsec) 
{\it HST}-STIS images was fixed by smoothing the STIS
image to the ground-based $I$-band resolution, matching all
sources $>5\sigma$, 
and then transforming coordinate grids using the IRAF task, {\it GEOTRAN}.
After maximizing the cross-correlation signal
between frames,  the match between $I<25$ optical 
sources has $\pm0.1$\arcsec\ rms.
Since the large $I$-band mosaic-CCD images are precisely aligned 
($\sim$0.3\arcsec\ rms) to the radio grid (e.g., Richards et al.\ 1999),
the HST images should be aligned to the radio frame to
$\sim$0.3\arcsec\ rms.

While ground-based optical imaging with $\sim1$\arcsec\ seeing
typically identified
only hints of extended sources, our STIS imaging
uncovers complex morphologies and 
distinct components with intervening low surface-brightness (SB) emission.
While only two of our sources currently have spectroscopic redshifts,
there are now over 50 robust measurements of field radio--submm 
sources with $<z>=2.4$ and an interquartile range of $1.8<z<2.8$
(Chapman et al.\ 2003b; S.Chapman in preparation), 
suggesting that our {\it HST} targets should
sample a similar redshift range. These measurements
have confirmed that the submm/radio redshift indicator
(Carilli \& Yun 1999) has large ($\sim$50\% 
rms) error, negating their use for studying optical luminosities. 
However, the angular diameter distance varies slowly over the redshift
range $z=1$--4, and the Carilli \& Yun indicator is sufficient to 
assume a physical scale for their angular diameters
(the submm/radio ratio indicates redshifts for our
sources lying between $z=1.3-3.1$ -- Fig.~2). 
All calculations assume a flat, 
$\Lambda$CDM cosmology with $\Omega_\Lambda=0.7$ and
$H_0=65$\,km\,s$^{-1}$\,Mpc$^{-1}$, so that 1\,arcsec
corresponds to 8.6\,kpc at $z=1$, 9.0\,kpc at $z=2$, 8.3\,kpc at $z=3$, and
7.5\,kpc at $z=4$.

\subsection{Archival images of LBGs and NB-galaxies}
In order to construct a morphological comparison sample of high redshift 
galaxies, we obtained {\it HST}-WFPC2 images from the {\it Canadian
Astrophysics Data Center} (CADC).
We concentrate on 
Lyman-break galaxies (LBGs -- Steidel et al.\ 2003):
rest-frame UV selected, star-forming
galaxies at $z\sim2.5-3.5$.
We also study Ly$\alpha$ emitters at $z\sim3.1$,
detected using a 80\,Angstrom narrow-band filter (similar to that presented in 
Steidel et al.\ 2000), which we denote NB-galaxies.

WFPC2 images in the F814W filter were obtained for literature 
Lyman-break galaxy (LBG) fields with at least 3 orbits of integration.
These fields include the 0000-263, 0347-383 fields presented originally
in Giavalisco et al.\ (1996), and the SSA22 field from Steidel et al.\ (1998).
In addition, parallel WFPC2 exposures from our own primary 
STIS program in the SSA22 field were used for this analysis.
All NB-galaxies lie in the SSA22 field.
The WFPC2 images have a pixel scale of 0.1\arcsec, and typically 
reach $I=25.5$ at 5$\sigma$, dependent on the precise exposure time in
the particular image.

We also use morphological parameters extracted  
for LBGs in the {\it Hubble Deep Field}, taken in the F606W filter.
Details of these measurements can be found in Conselice et al.\ (2003).

\section{Results}

\subsection{Submm galaxies: General morphology}
We begin our analysis of the submm galaxy morphologies with a general 
assessment of their properties.
Fig.~2 shows the radio-identified optical components to the submm
sources, the photometric redshift (or spectroscopic redshift when available),
and identifies the image with Table~1.
The images in Fig.~2 have been smoothed with contours overlaid
to increase the visibility of faint structures in the submm galaxies.
For an unsmoothed version of this figure, see the companion paper
(Conselice, Chapman \& Windhorst 2003).
Descriptions of the individual objects in the sample can be found in 
Appendix~1.

Inspection of the images in Fig.~2 reveals that many of the submm
galaxies display a distinct 
extended, linear morphology. A brighter knot
is often asymmetrically displaced toward the end of a linear, lower
surface brightness nebulosity surrounding the compact components.
Multiple components are often revealed within $\sim$1--2\arcsec\ scales.
A few of the submm galaxies look similar to {\it chain} galaxies (Cowie, Hu \&
Songaila 1995), while others are suggestive of 
mergers in progress, a statement which we will
further quantify below.



Local infrared-luminous galaxies have been identified with
merging galaxy morphologies (Sanders \& Mirabel 1996).
As a simple, illustrative comparison, 
the submm galaxy component separations and elongations
can be compared to the highest luminosity local galaxies: the
ULIRG subset from 
the 1 Jy sample of {\it IRAS} galaxies from Veilleux, Kim \& Sanders (2002
-- see also Murphy et al.\ 1996).
As shown in Fig.~3
the distribution in component separation of local ULIRGs 
is peaked at small values but has a significant tail at higher values.
These measurements consider components with nuclear magnitudes of $M_R < -20.1$.
The trend in Veilleux, Kim \& Sanders (2002) is for the highest
luminosity ULIRGs (those with $L_{ir}>3e12\,L_\odot$) to have significantly
smaller nuclear separations than the $1e12<L_{ir}<3e12$ ULIRGs.

We place the submm galaxies on the same plot by measuring the maximum extent
between multiple component structures, centroiding on the outermost intensity
peaks within coherent structures (those exhibiting low surface-brightness
bridges). The submm galaxies by contrast all have projected
separations of $>$5\,kpc, except for sources (7) and (8)
which do not exhibit multiple component or extended structures.
The submm galaxy STIS images likely represent rest frame wavelengths
of 1500 to 3000 Angstroms, and we must be cautious of direct comparison with 
the $R$-band local fiducial from Veilleux, Kim \& Sanders (2002).
Consideration of rest frame-UV images of ULIRGs (Goldader et al.\ 2002) 
suggests that UV-bright star clusters can become apparent within galactic
nuclei, but do not significantly affect the global separation measurements
presented in Fig.~3.
 
%
This result is suggestive that the merger configurations of high redshift
submm galaxies are typically larger and at 
an earlier stage than those of comparable 
luminosities at low redshift identified in the infrared by IRAS.

\subsection{Submm galaxies: quantitative morphology}

In order to associate quantitative measures to these visual impressions of
the submm galaxies, we undertook 
a morphological analysis of the STIS images of submm galaxies 
under the {\cal LMORPHO} environment (Odewahn et al.\ 2002).  
Catalogs of all sources in each of the
9 STIS fields were first prepared, covering an area
0.00161~deg$^2$. 
An initial SExtractor (Bertin \& Arnouts 1996) catalog was assembled,
computed with a 1$\sigma$ threshold, a fixed 5-pixel aperture
and a pre-detection smoothing Gaussian with $\sigma=5$ pixels in a $9\times9$
pixel filter. A total of 619 sources were cataloged. 
An interactive 
editing tool was then used to perform star-galaxy separation and 
source-detection
cleaning (i.e., fixing obvious mistakes in image segmentation by SExtractor
or eliminating detected image defects, etc.), yielding
611 sources 
with photometry from the automated galaxy surface photometry
package ({\it GALPHOT}). 
The submm sources, along with the 102 galaxies with nearby companions
($<1$\arcsec), were isolated, and
{\it GALPHOT} was re-run interactively to  
locate and eliminate
nearby sources in the final photometry, and to determine 
more reliable local sky values to improve the photometry.

The submm optical counterparts are detected with sufficient S/N ($>10$ per
beam) for detailed morphological analysis, but
are clearly much too faint for application of a Fourier-based classifier
(e.g., Odewahn et al.\ 2002).
We estimate standard quantitative measures of galaxy morphology.
Table~1 lists ground-based submm, radio, and I-mag photometry, along with
these morphological parameters extracted from the {\it HST} images:
total magnitude, effective radius - $R_{50}$, concentration index,
aspect ratio, image size, and mean surface brightness.

The {\cal CAS} (concentration, asymmetry and
clumpiness) program for morphological analysis (Conselice 2003)
was also employed to quantify the asymmetries of the submm galaxies.
{\cal CAS} is
based on the idea that structural and morphological features of galaxies
are directly related to past and present underlying physical processes.
The processes traced by {\cal CAS} are the past and present
star formation and merger activity.  In the {\cal CAS}
system, the asymmetry index ($A$) is used to determine whether or not
a galaxy is involved in a major merger (Conselice et al. 2000; Conselice
2003; Conselice et al. 2003).  In this system major mergers are
always found to have an asymmetry greater than some limit, which in
the rest-frame optical is $A_{\rm merger}$ = 0.35.  The concentration index
is also a fair representation of the scale of a galaxy and is proportional
to the fraction of stars in a bulge component (e.g., Graham et al. 2001;
Conselice 2003). 

In a similar manner to the {\cal LMORPHO} analysis, galaxies with nearby
companions ($<1$\arcsec), were isolated before calculating morphological
quantities.
We placed our initial guess for the center on the brightest portion of
the sub-mm galaxy, and then run the CAS program to determine the
asymmetry and light concentrations for these galaxies.  This is done
through well defined radii, centering and background removal methods
which are fully described in Conselice (2003).  
Morphological parameters extracted include asymmetry, concentration,
and a growth-curve estimate of effective radius, or Petrosian radius
(Bershady, Jangren \& Conselice 2000; Conselice, Bershady \& Jangren 2000),
and are listed in Table~2.
The Petrosian radius (in units of arcsecs) 
is defined as $1.5\times r_{(\eta = 0.2)}$, where $\eta = 0.2$
is the radius ($r$) where the surface brightness within an annulus at $r$ is
1/5 the surface brightness within $r$.
This provides a measure of size which
doesn't depend on isophotes, complementary to that of the  
image moment-based $R_{50}$ from the {\cal LMORPHO} analysis above.
Corrections have been applied
to the asymmetry and concentration values
using an offset computed through the simulation how
local normal galaxies would appear at 
the redshifts of submm galaxies (Conselice, Chapman \& Windhorst 2003
-- a companion paper which provides
detailed {\cal CAS} analysis of the submm galaxies and their merger fraction
relative to other galaxy populations).

\subsection{Comparison of the submm galaxies with the field catalog}

Morphological analysis of faint sources can be susceptible to differences
in the image depth, instrumental response and pixel sampling
(e.g., Odewahn et al.\ 2002).  Therefore, 
the most direct approach to studying the morphologies of the submm sources
is to compare the photometric properties
of the submm/optical counterparts with the general population of 
optical sources in the same 9 STIS frames. 
The results of this analysis are presented in Fig.~4.

Fig.~4a plots the isophotal effective radius ($R_{50}$)
as a function of total magnitude. 
$R_{50}$ is a measure of the deprojected radius containing
50\% of the light represented by the total magnitude.
Most of the submm sources (8 of 11) 
exhibit larger effective radii ($R_{50}>0.35$) per magnitude interval
than the general population.
In the STIS-magnitude bin 25--27 (subsuming the submm galaxies)
the median $R_{50}=0.18$, with an interquartile range 0.12--0.23.
We emphasize that while sources with $z<1$ cannot be compared
directly with the submm galaxies, they must exhibit smaller physical scales
per unit angle, making the comparison with the extended
submm galaxies even more dramatic. 
Before attempting to interpret this result,
it is important to consider how $R_{50}$ is calculated. 
Each pixel is deprojected based on the isophotal ellipse shape and
orientation in the usual manner, giving a measure of the pixel
distance from center in the equatorial plane of the galaxy. These
spatial measures are binned radially and used to compute the
mean surface brightness. Integrating and extrapolating this
profile produces a total magnitude and a growth curve.
Using this growth curve we derive $R_{50}$, a measure of
the deprojected radius at which 50\% of the total light is collected.
As galaxies become more edge-on, there are fewer sampled
points to use for the deprojection in the semi-major axial direction.
An infinitely  thin, perfectly edge-on galaxy cannot have its
$R_{50}$ calculated in this manner. As a result, galaxies
approaching this idealization can exhibit $R_{50}$ values
larger than physically represented. This systematic effect,
combined with the highly elongated nature of our sub-mm galaxy
images, causes the submm galaxy points in Fig.~4a to occupy the
upper $R_{50}$ envelope of the point distribution that is
dominated primarily by {\it roundish} field galaxies.

The {\cal CAS} estimate of the Petrosian radius provides a complementary
analysis to the $R_{50}$ result.
The average Petrosian radius of the submm galaxies (2.1\arcsec), is larger
than any other star-forming galaxies at low or high redshift
(after taking redshift effects into account -- Conselice et al.\ 2000).
Lyman-break galaxies from the {\it Hubble Deep Field} have a Petrosian radius 
of 1.2\arcsec (Conselice et al.\ 2003).
The {\cal CAS} analysis thus supports our finding using the $R_{50}$
index, that submm galaxies are larger systems than other star-forming galaxies.
We have also verified our finding from the pair separation

Analysis of the field galaxies exhibiting a similar range of 
$R_{50}$ to the submm galaxies reveals that they are split
approximately in half between apparently large galaxies (likely to be 
nearby) and linear/edge-on galaxies. The latter are reminiscent of the
submm galaxy morphologies. 
There are 12 field galaxies within the magnitude range $R'(573)=$24--27 
(subsuming the magnitude range of the 8 submm galaxies with large $R_{50}$) 
having morphologies {\it similar} to those of the submm galaxies.
As we have targeted our STIS images around relatively bright ($>$5\,mJy)
and rare (0.25\,arcmin$^{-2}$) submm sources, we must scale by this
value to calculate the total number of field galaxies with similar
morphologies.
This suggests that there are roughly 4$\times$ the number of field galaxies
with similar magnitudes and morphologies to the submm galaxies,
but lacking the copious bolometric luminosities 
(the field galaxies are undetected in the submm and radio images
covering the same HST-STIS images). 

 
Fig.~4b shows concentration index ($C32$ = ratio of 75\% to 25\% 
quartile sizes) versus magnitude. 
Higher $C32$ implies a more compact light distribution
(i.e., stars have large $C32$). We note that submm
sources follow an apparently opposite trend compared to the control sample.
The mean $C32$ stays constant or increases slightly with fainter mag for the
submm sources, but decreases slowly for the control galaxy sample. 
For the general population, 
we are simply seeing the effect of decreasing S/N and effective resolution as
magnitudes become fainter: everything becomes diffuse (lower $C32$).
The faint ($R'(573)>26.5$) submm sources 
have a higher $C32$ than the control sample,
suggesting submm selection picks out more concentrated galaxies.

Finally Fig.~4c shows the image size (in arcsec) derived from the 
intensity-weighted image moment on the abscissa, 
and the mean surface-brightness (SB -- mag per arcmin$^2$) 
within the effective radius on the ordinate.
The data are discrete on the abscissa as image sizes
are established in integral pixel units. 
The submm sources are average or fainter in mean SB per size interval
than the general population. 
We demonstrated above that submm galaxies are more extended than the field 
galaxies (bigger $R_{50}$ and Petrosian radius, and larger component
separations than local infrared-luminous galaxies).
The large, multi-component configurations tend to low global SB measurements.
Dust extinction, expected to be significant in submm galaxies, will also make 
the rest-UV emission fainter. 
Surface brightness dimming at the high redshifts of the submm galaxies
$\propto (1+z)^4$ would also contribute to this trend.

While higher surface brightness core components are often present
in the submm galaxies,
the average surface brightnesses (Fig.~4c)
are not unusually high compared to the
field STIS population. At $\sim$25\,mag\,arcsec$^{-2}$ and
redshifts with a median $z\sim2.4$ (Chapman et al.\ 2003a),
they correspond to about
ten times that observed in a typical spiral-disk locally.

\section{Comparison with Lyman-break galaxy morphologies}

Lyman-break galaxies (LBGs -- Steidel et al.\ 1996, 1999) 
represent rest-frame UV selected, star-forming
galaxies at similar redshifts to submm galaxies ($z\sim2.5-3.5$). 
LBGs have HST morphologies which often 
exhibit distorted and irregular morphologies
(Giavalisco et al.\ 1996, Steidel et al.\ 1996, Erb et al.\ 2003).
However, there has never been a chance to compare against the submm-selected
galaxies.
In Fig.~5, we have reproduced a complete sample of
LBGs from Steidel et al.\ (2003) 
using sources from the fields presented in Giavalisco et al.\ (1996)
and Steidel et al.\ (2000), as described in \S~2.1, with the goal of 
applying the same analyses as were performed on the submm galaxies.

Another class of high redshift galaxies are the Ly$\alpha$ emitters
(e.g., Cowie \& Hu 1998).
These galaxies have been isolated at $z=3.1$ using a 80\,Angstrom narrow-band
filter and compared to the LBG population in the same field (Steidel et al.\
2000). We have recently obtained a much larger image of the same region
in a similar filter (S.~Chapman, in preparation).
Steidel et al.\ (2000) have demonstrated that only 25\% of LBGs satisfy the
large equivalent widths in Ly$\alpha$ to be detectable in such narrow-band 
(NB) images.
In Fig.~6 we show the HST images of all NB-galaxies in the SSA22 
field, lying
within the WFPC2 archival 
images. As these objects are generally much fainter than the
LBGs, we have smoothed the images for visibility.

\subsection{Morphological parameters}
To compare the LBGs and NB-galaxies to the submm galaxies in a self-consistent
manner, we applied the same morphological analysis to the complete galaxy
catalogs from the WFPC2 images using the same {\cal LMORPHO} environment.
This enables the same normalization against the field population as we 
did for the STIS images of submm galaxies (\S~3.2), removing any instrumental
biases. 
The effective radius in particular (Fig.~4a) suffers from large
instrumental dependency due to the pixel scales of the images.

Fig.~4\,a\&b suggest that neither LBGs nor NB-galaxies 
generally differentiate
themselves from the field galaxy population, in either effective radius
or concentration index. This is opposite to the finding from \S~3.2
for the submm galaxies which show larger $R_{50}$ 
and an opposite trend in concentration index than the field galaxies.
There do appear to be several LBGs at the fainter end of the total magnitude
scale which are identified as having larger effective radii.
Inspection of these sources reveal them to be amongst the closest 
morphological matches to the submm galaxies.
We emphasize that the submm galaxies and LBGs can only
be compared explicitly to their respective field samples, and not to each
other, due to the sensitivity of these parameters to the STIS and WFPC2
instruments.

Fig.~4c reveal the LBGs to have surface-brightnesses $\sim1.3$-mag 
higher than the submm galaxies. The NB-galaxies are comparable in 
surface-brightness to the submm galaxies.
This reflects the trend for LBGs to be dominated by a central knot
of high-surface-brightness surrounded by lower surface-brightness
nebulosity, as described previously in Steidel et al.\ (1996). 

We can also study the {\cal CAS} parameters of the submm galaxies
compared to LBGs (here the NB-galaxies are not considered due to typically 
lower S/N).
As discussed above, and in Conselice, Chapman \& Windhorst (2003), 
the {\cal C,A} values for the submm galaxies and LBGs
have been corrected by redshifting nearby galaxies to $z=3$ and
seeing how the average {\cal C,A} values change due to redshift effects.
As mentioned above, the average size of the submm galaxies are larger than 
the LBGs: 2.1\arcsec\ (submm) versus 1.2\arcsec\ (LBGs).
The submm galaxies also distinguish themselves from the LBGs
in terms of concentration, but are similar in asymmetry. 
Fig.~7
plots {\cal A} and {\cal C} for both LBGs and submm galaxies
where the submm galaxies have about the same asymmetry in the median
but are more concentrated (3.2 versus 2.1 for LBGs).
This difference may reflect the extended, bright starbursts or AGN in the 
submm galaxies.

\subsection{A qualitative classification scheme} 

Traditional quantitative measures of morphology have difficulty contending
with the range of distorted, merger configurations exhibited by the
submm galaxy population. Subtle features of the morphologies can be missed
by applying algorithmic processes to the inherently irregular light 
distributions.
To complement our constraints on the quantitative morphological comparison
of high redshift galaxies, 
we devise a simple classification scheme to encompass the range of submm galaxy
morphologies, which distinguishes the main
features seen in our HST images.
The categories and memberships are listed in Table~3.

We delineate the following five categories:
compact regular, compact irregular, elongated regular, elongated irregular,
multiple component/elongated irregular.
These categories were applied independently to the submm galaxies by
members of our group, with the same results.
The majority (83\%) of the submm
galaxies are clearly in the latter two categories,
with combinations of irregular, multiple components.
Only two submm galaxies (sources 7\&8) can be considered isolated and
compact (compact regular category).

We attempt to place the LBGs and NB-galaxies within the same
qualitative classification scheme that we applied to the submm galaxies.
We have used both the unsmoothed and smoothed images to study the morphologies
for consistency with the submm galaxies, and in order not to miss faint
extended structures.

One difficulty with the comparison is that the LBGs are typically optically
brighter than the submm galaxies, displaying high central surface-brightness
cores surrounded by lower surface-brightness nebulosity (Steidel et al.\ 1996;
Giavalisco et al.\ 1996; Fig.~5).
The submm galaxies are generally lower surface-brightness objects, with only
bright cores rising above the noise level.
However, we would expect this difference to produce a one-way bias, in the
sense that the submm galaxies should only appear less irregular due to loss
in surface-brightness sensitivity.
By contrast, the NB-galaxies are sometimes even fainter than many of 
the submm galaxies and even more structure may be lost in the noise.

The placement of the various LBGs into the categories outlined for the 
submm galaxies is often challenging, whereby very faint sources neighboring
more dominant central sources are often present.
While these could be classified as `multiple component', they do not appear
to represent the same type of major merger configuration as the
submm galaxies (see also Conselice, Chapman \& Windhorst 2003).
We have typically classified these sources as compact irregular, 
when the separations of the peaks are less
than 0.5\arcsec.

Compared with the LBGs, the submm galaxies typically appear more extended
with multiple components of larger separation, although both
populations exhibit similar morphologies in a few of their representatives.
The NB-galaxies are more difficult to classify, but also generally appear to 
be more compact than the submm galaxies, although less so than the LBGs.
The summary of this comparison is listed in Table~3.

\section{Discussion}

Our {\it HST}  images have identified the 
counterparts of
submm galaxies and allowed a quantification of their morphological 
properties.
The galaxies are characterized by larger sizes than the field
population in the same STIS images, confirmed by $R_{50}$ and
growth curve estimates of
the radii (Petrosian radii) compared to other populations.
The isophotal $R_{50}$ 
values provide a 
quantitative index which separates the peculiar morphologies of the submm
galaxies from the bulk of the field population.
It also appears to differentiate the submm galaxies from other high-redshift
star-forming galaxies, the LBGs and NB-galaxies, which generally do not
distinguish themselves from the field population in the basic 
morphological parameters.

However, the $R_{50}$ index does not fully describe the stunning
morphologies of the submm galaxies.
Many show multiple components or extended
structure beyond that extracted by the analysis routines.
The morphologies appear to extend over scales conceivably up to 5\arcsec\
($\sim$40\,kpc), including regions beyond the radio emission, and 
therefore unlikely 
to be directly emitting the bulk of the bolometric luminosity.
The separations of components in the submm galaxies are typically larger
than comparable luminosity systems in the local Universe (Fig.~3).
These morphologies are contrasted with the appearances of LBGs and 
NB-galaxies, which are generally more compact. 
The comparably large asymmetries of submm galaxies and LBGs relative to 
local normal galaxies 
indicates that LBGs are still highly irregular systems, although typically
smaller. 

The submm galaxies typically have implied SFRs of 
$\sim$1000\,M$_\odot$\,yr$^{-1}$, and total projected areal coverage 
of $\sim$50\,kpc$^2$ traced in the rest-frame UV light
from young stars observed by our STIS imagery
(assuming the typical redshifts of the submm galaxies
-- Chapman et al.\ 2003a).
This corresponds to a SFR density of 
$\sim$20\,M$_\odot$\,yr$^{-1}$\,kpc$^2$.
Lehnert \& Heckman (1996) have suggested that local star-forming galaxies
are self-regulated by a {\it maximal starburst} of
the same $\sim$20\,M$_\odot$\,yr$^{-1}$\,kpc$^2$ (for a Salpeter IMF).
While the equality of the numbers is likely coincidental, the large
sizes of the submm galaxies are suggestive that starbursts of this
order ($\sim1000$\,M$_\odot$\,yr$^{-1}$) cannot exist in smaller galaxies.
Contrast the smaller sizes and lower characteristic SFRs of the LBG population.

There is, however, no way to conclude from the present data whether an AGN
may be responsible for some fraction of the bolometric luminosity.
The resolved nature of submm galaxy morphologies suggests that
strong optical QSO components are not common in the submm population.
The large UV-extent of the submm galaxies may be evidence that
star formation dominates the dust heating in many cases.
Indeed, the only isolated source in our sample (SMM\,J123713.9) is
also the smallest and most compact, and
is in fact the brightest Chandra X-ray source
from our radio-identified submm population in the HDF region,
suggesting a dominant AGN (see also Alexander et al.\ 2003).
This hypothesis is bolstered by the high spatial resolution radio measurements
using the MERLIN radio interferometer ($\sim$0.3\arcsec\ synthesized beam)
of several sources from our HST sample (T.\ Muxlow, in preparation). 
In $\sim$50\% of the cases, the radio emission is extended, and traces the
UV morphology.

Can mergers truly produce a linear, elongated
structure, reminiscent in some cases of
{\it chain} galaxies (Cowie, Hu \& Songaila 1995)?
To address this question we have studied the
fraction of low-redshift ULIRG merger systems which
look like linear, elongated structures.
As described in Conselice, Chapman \& Windhorst (2003)
HST images of low-$z$ ULIRGs are redshifted to the submm galaxy distances.
14 out of 51 (27\%) ULIRGs studied in this manner have
structures that look like {\it chain}
galaxies. This is a comparable fraction to the
submm galaxy morphologies which also look like {\it chain} galaxies.
Comparing with
Cowie, Hu \& Songaila (1996), the fraction could be argued to be even higher.
We therefore consider $\sim30$\% to be a conservative limit of the
number of low-$z$ ULIRGs which have morphologies consistent with the submm
galaxies.

The components of submm galaxies may 
therefore represent the first generation of merging of substantial
fragments of galaxies.
The large fraction of highly elongated or
linear structures (some similar to the {\it chain} galaxies of
Cowie, Hu \& Songaila 1995) are suggestive of early
stages of a dynamical event where two approximately equal mass clumps have
passed by or through each other. If the components do not subsequently reach
escape velocity, they will fall back into each other and become a
merger in the near future. Such an initial dynamical event would
induce star formation on the dynamical time scale for the system.
Submm galaxies may be hosted by very high mass halos, based
on their strong redshift clustering (Blain et al.\ 2004). This is consistent
with the CO molecular gas emission line
widths and possible rotation curves
(Frayer et al.\ 1998, 1999; Genzel et al.\ 2002; Neri et al.\ 2003).
If we assume only the current maximum projected separations for the submm
galaxies (Fig.~3), and enclosed dynamical masses
of order $\sim$5$\times$10$^{11}$\,M$_\odot$,
we can use Kepler's third law to calculate relaxation timescales.
The median relaxation time is 30\,Myr with an interquartile range of 31\,Myr.
Hydrodynamical models have shown 
that SNe may drive out the dust more easily than the gas in lower-mass 
star-forming galaxies (Mac~Low \& Ferrara 1998).
LBGs may be longer-lived and lower-mass galaxies which expelled the
bulk of their dust prior to experiencing a luminous event of the
submm galaxy class.


In an accompanying paper, Conselice, Chapman \& Windhorst (2003) demonstrate
that up to 80\% of the submm galaxies 
have morphologies consistent with major mergers.
However, the radio emission typically
points us to a single optical source, and only in the case of
SMM\,J141809.8 (Chapman et al.\ 2002a) do we have spectroscopic confirmation
that three components lie at the same redshift ($z=2.99$).
In addition,
high redshift galaxies often have complex morphologies
(e.g., Cowie, Hu \& Songaila 1995), and many appear qualitatively similar
to the extended knotty structures shown in Fig.~2 (although we have
already considered in detail the LBG and NB-galaxy morphologies, finding
smaller and less disturbed configurations than the submm galaxies). 
Alternative explanations to mergers have been presented in the literature.
They could be associated with
planar structures in the galaxy formation process, or they could
be structures generated by sequential star formation 
(Cowie, Hu \& Songaila 1995).

Can we decide if an apparent clump of components, seen in many of our sources 
in Fig.~2, is truly a merger versus an assembly of 
HII regions in a large galaxy?
We previously showed how SMMJ141809.8 displays a striking difference 
between its rest-frame UV and Visible emission, 
a {\it morphological K-correction},
which is seen only rarely in more local galaxies (Hibbard \& Vacca 1997;
Abraham et al.\ 1999; Kuchinski et al.\ 2001; Windhorst et al.\ 2002).
However the most infrared-luminous local galaxies often show different
morphologies in rest-frame nearIR (Scoville et al.\ 2000;
Dinh-V-Trung et al.\ 2001).
We must question how morphological K-correction might affect our 
interpretation of the submm galaxies.
We must also address how much structure is being resolved out by HST, lost
to surface brightness dimming at redshifts $z>>1$.
A well-studied local ULIRG, Mkn231 (Goldader et al.\ 2001),
shows 75\,kpc tidal tails (10\arcsec\ at the submm galaxy redshifts).
While we expect to miss many such features from the strong
surface brightness dimming effects at high-$z$, we note that many
of the optical morphologies do exhibit signs of extension.
Colley et al.~(1996) first made the argument
that faint clumps at the same $z$ close on the sky may be part of a bigger
galaxy, most of the underlying part not being visible due to SB-dimming (the
knots are unresolved, so they only dim as $\propto (1+z)^2$).
Without redshifts and velocity information for individual
components, we cannot claim any source is a merger in progress.
However, if it is not a merger yet, then it will likely be a
future merger when the pieces come together and violently relax (barring the
unlikely situation that $\sigma_v$ is large enough for the violent relaxation
to never happen).
Regardless of the interpretation, these objects are clearly star formation
wrecks of some sort,
consistent with the submm excess and plausibly young ages.

\section{Conclusions}
The HST-STIS images of submm galaxies 
have clearly identified many large, distorted,
and plausibly multiple component, merger systems. 
The physical separations of components are larger than similar luminosity
IRAS galaxies seen locally, suggestive that they may typically represent
an earlier stage of the merger process.
The faint ($R'(573)>24$) submm sources on average 
have a larger effective radius ($R_{50}$)
than the general ``field'' sample defined within the same STIS images.
They are also typically larger than the $z\sim3$ LBG population,
defined in terms of $R_{50}$ or Petrosian radius.
The rest frame UV extents and far-infrared estimated SFRs 
of these submm galaxies are consistent with
predictions based on the {\it maximal} star formation rate density of
20\,M$_\odot$\,kpc$^{-2}$ seen in local starburst galaxies.

Consideration of the field sources with comparable $R_{50}$ values 
often reveals highly elongated and distorted systems, some indistinguishable
from the morphologies represented by the submm galaxies. Scaling 
by the source count of the submm galaxies at $>$5\,mJy (0.25\,arcmin$^{-2}$)
suggests that there are roughly 4$\times$ the number of field galaxies 
with similar magnitudes and morphologies to the submm galaxies which do not
obviously generate the copious bolometric luminosities 
(they are undetected in the submm and radio images).
Our analysis of the LBGs and NB-galaxies suggests that some of these field 
sources with morphologies similar to submm galaxies would be LBGs/NB-galaxies.
It remains to be seen whether the bolometric luminosities and SFRs of
the largest, irregular galaxies in the field are generally larger than the
other field galaxy populations.


\acknowledgements
We would like to thank Ian Smail and Andrew Blain for detailed suggestions 
and comments on the manuscript which greatly helped improve the paper.
We also acknowledge the detailed suggestions of an anonymous referee.
Based on observations made with the NASA/ESA Hubble Space Telescope, obtained 
[from the Data Archive] at the Space Telescope Science Institute, 
which is operated by the Association of Universities for Research in Astronomy,
Inc., under NASA contract NAS 5-26555. These observations are associated
with proposal \#9174.
Support for proposal \#9174 (SCC, RW) was provided by NASA through a grant from
the Space Telescope Science Institute, which
is operated by the Association of Universities for Research in 
Astronomy, Inc., under NASA contract NAS 5-26555.

\clearpage
\appendix
\centerline{\bf Notes on Individual objects}
\bigskip
{\bf (1) SMM\,J123553.3+621338}  
A very extended galaxy with as many as 10 compact components
arranged in a north-south assembly within 3.4\arcsec.
Submm flux (Chapman et al.\ 2001b), radio flux (Richards 2000).

{\bf (2) SMM\,J123600.1+620254} 
Five knots along a curve indicate that
this galaxy is possibly an edge-on merger.
Submm flux (Chapman et al.\ 2001b), radio flux (Richards 2000).
   
{\bf (3) SMM\,J123616.2+621514} 
A trio of components suggest a system about to merge.
To within the astrometric error,
the brightest component appears to be
aligned with the radio emission, a Chandra X-ray source (both hard and soft
bands), and a $K$-band peak in the ground based (0.5\arcsec) image.
Submm flux (Chapman et al.\ 2001b), radio flux (Richards 2000).

{\bf (4) SMM\,J123618.3+621551} 
While only the central source is radio identified, the additional clumps
lying within $2$\arcsec\ are plausibly associated, suggesting
an early stage merger.
Submm flux (Barger, Cowie, \& Richards 2000), radio flux (Richards 2000).

{\bf (5) SMM\,J123621.3+621708} 
A strong radio peak with an extremely faint optical identification
lies offset 2\arcsec\ from a weaker radio peak associated with a
north-south linear structure.
Submm flux (Barger, Cowie, \& Richards 2000), radio flux (Richards 2000).

{\bf (6) SMM\,J123622.7+621630} 
The radio centroid sits in a saddle of low surface-brightness emission
amongst a spectacular edge-on merger exhibiting 5 brightness peaks within an
extended linear structure. The source is detected by Chandra in the hard
X-ray band, but not in the soft band.
Submm flux (Barger, Cowie, \& Richards 2000), radio flux (Richards 2000).

{\bf (7) SMM\,J123710.0+622649} 
This source has two radio sources lying within the SCUBA beam. The western
source corresponds to a bright elliptical galaxy, the redshift of which is
unknown. A photometric redshift from the available multiband imaging
($B, V, I, K$) suggests a $z\sim0.4$ galaxy.
There is no significant optical
emission at the location of the eastern radio source.
Submm flux (Chapman et al.\ 2001b), radio flux (Richards 2000).

{\bf (8) SMM\,J123713.9+621827} 
The radio peak is offset by 0.7\arcsec\ from an isolated and compact
optical source, showing some extent in the north-east.
This source is a bright Chandra X-ray source, and is the strongest
submm emitter in our sample (16~mJy). With a radio flux of 595\,mJy, it
is also the strongest radio emitter.
This source is likely to be a heavily dust obscured AGN.
Submm flux (Chapman et al.\ 2001b), radio flux (Richards 2000).

{\bf (9) SMM\,J131231.9+424430} 
A double extended source, with signs of low surface brightness
features. This galaxy was detected at 6.7\mum\ using the ISO satellite
by Sato et al.\ (2002), suggesting that many of these submm galaxies will
be routinely detected by the SIRTF mission.
Submm flux (Barger, Cowie \& Sanders 1999), radio flux (E.\ Richards, private
communication).

{\bf (10) SMM\,J131235.2+424424} 
This object is almost undetected, with only a faint $1.9\sigma$ peak
rising above the noise at the radio position (omitted from Fig.~2).
As the source was detected at $I=26.4$ in ground based imagery, the HST
may be resolving out a complex and diffuse structure.
Submm flux derived from our own reduction of the archival SCUBA data,
radio flux (E.\ Richards, private  
communication). The submm and radio fluxes are presented in the figures
of Chapman et al.\ (2002c).

{\bf (11) SMM\,J141809.8+522205} 
An assembly of bright blobs, with low surface brightness intervening material.
A bridge of emission connects the strongest two components.
Spectroscopic redshifts have been obtained for all three central
blobs, $z=2.99$ (Chapman et al.\ 2002a -- see this work also for radio
and submm flux measurements). 
The original radio data in this region was presented in 
Fomalont et al.\ (1991).
The UV spectroscopic redshift for the 
brightest $R$-band component is presented in Chapman et al.\ (2000).
A strong $K$-band source lines up with northernmost blob,
coincident with a millimeter interferometry measurement

{\bf (12) SMM\,J221724.7+001242} 
A filamentary object, which appears
like an irregular edge-on galaxy, flanked by two compact
components. The galaxy is identified in the VLA B-array radio map
and in the archival Chandra X-ray image.
Lensing from the bright
galaxy to the north-west may be subtley distorting the image
(Chapman et al.\ 2002b). Fitting and subtracting the bright elliptical
galaxy does not reveal any additional structure or components.
The updated radio and submm fluxes for this source are presented in
Chapman et al., (in preparation).

{\bf (13) SMM\,J221726.1+001239} 
This
is the ``blob-1'' submm source
(Steidel et al.\ 2000; Chapman et al.\ 2001b)
lying at the center of a $z=3.09$ proto-cluster.
The HST image reveals a faint,
apparently linear system, with surrounding very faint components
which could represent a merger in progress.
A brighter unresolved core along the linear structure suggests
an AGN or a more concentrated starburst.
While no significant radio emission has been detected to isolate the
the position of the submm source,
an interferometric detection of molecular gas in CO(4-3)
identifies the submm emission with this HST source 
(Chapman et al.\ 2004).
The radio and submm fluxes, and the spectroscopic redshift
 for this source are also detailed in (Chapman et al.\ 2004).


\begin{figure*}[htb]
\centerline{
\psfig{figure=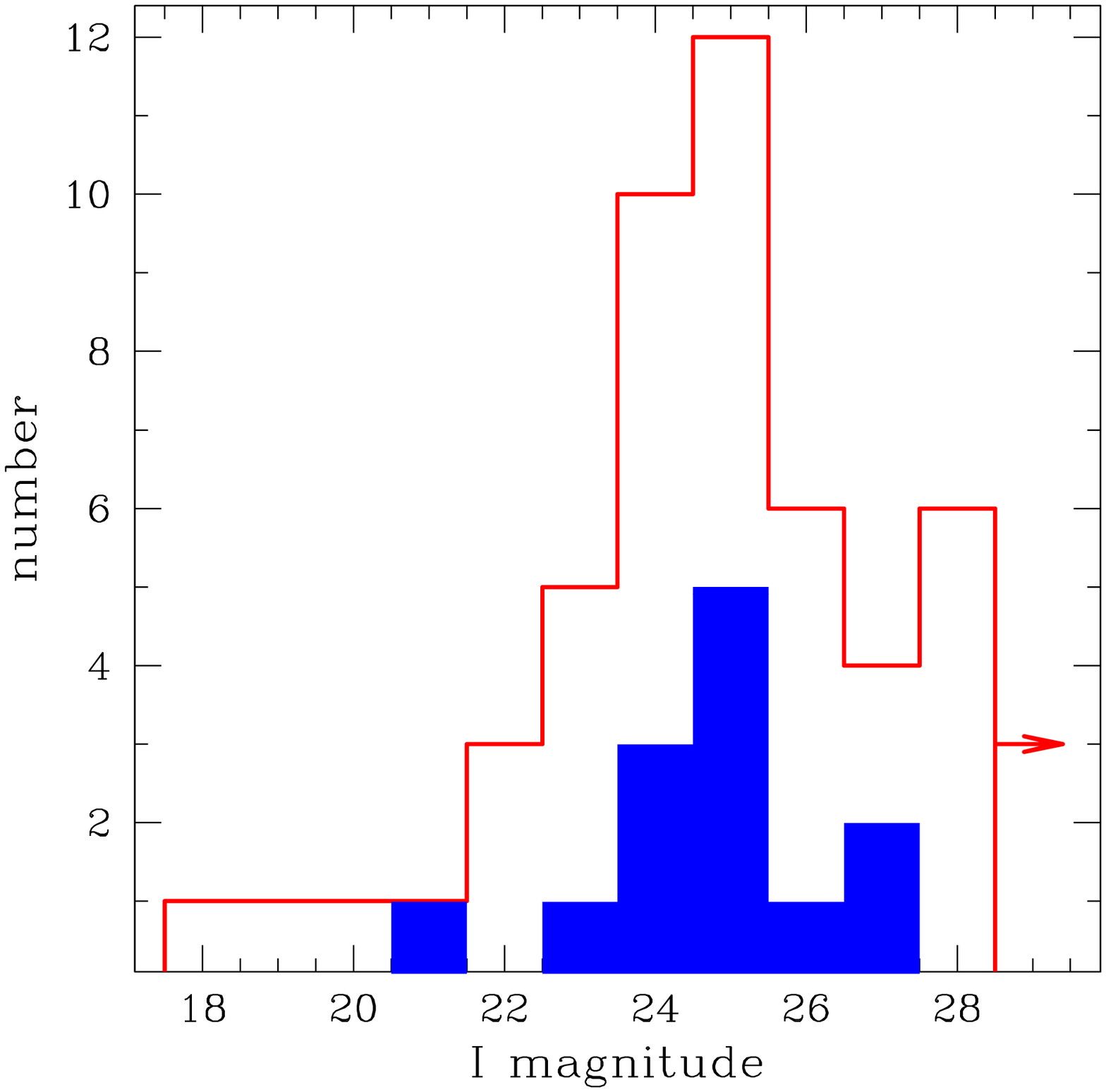,angle=0,width=5.8in}
}
\caption{\footnotesize
$I$-band magnitude distributions of the parent sample of submm galaxies
identified through their radio distribution (line histogram -- from
Chapman et al.\ 2003b), compared with 
the $I$-mags of the HST sample considered here (shaded histogram).
The faintest bin of the parent sample represents mostly lower limits to the
$I$-band flux.
The HST sources appear representative of the parent distribution.
}
\label{fig0}
\end{figure*}

\begin{figure*}[htb]
\centerline{
\psfig{figure=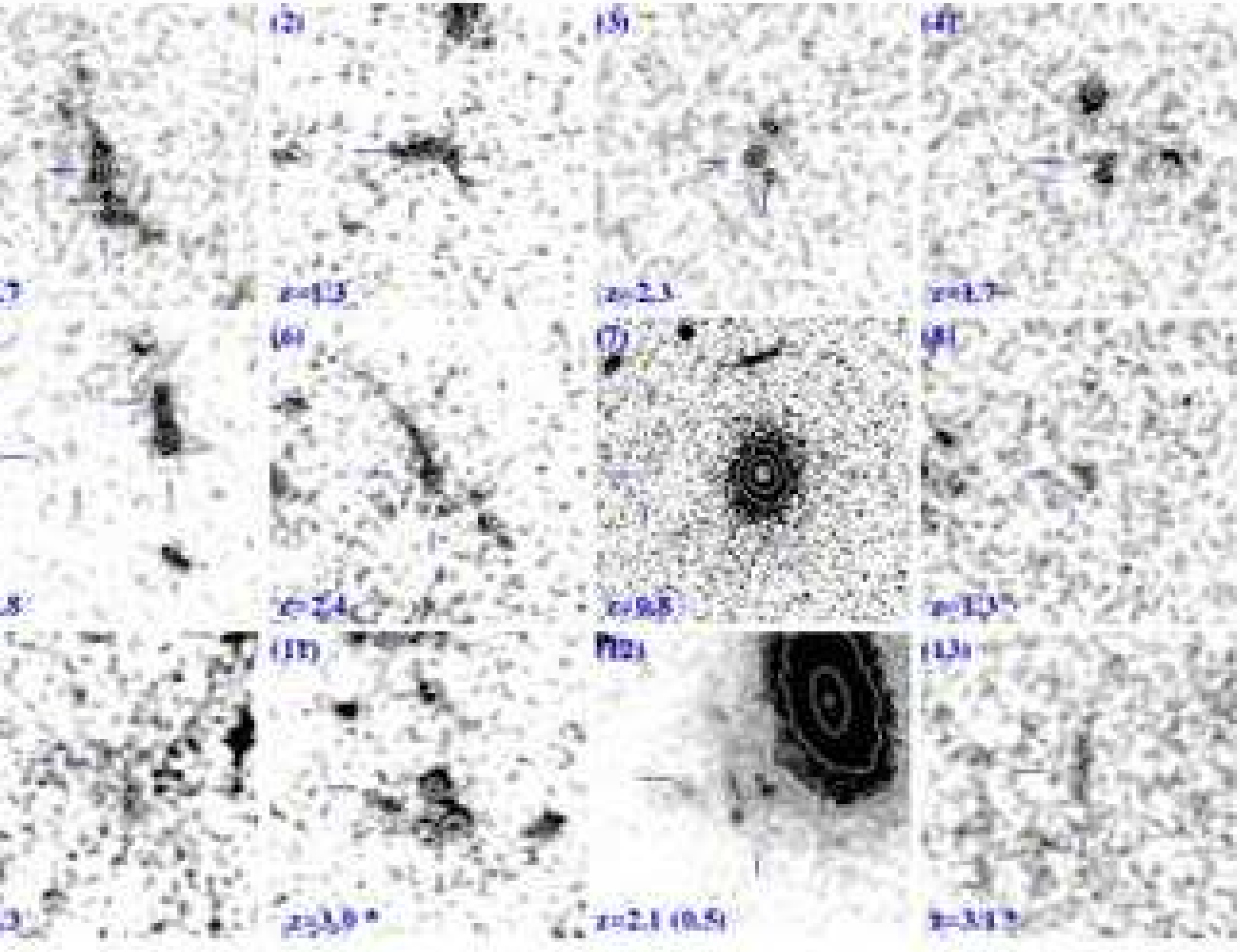,angle=-90,width=8.8in}
}
\caption{\footnotesize 
{\it HST-STIS} observations of submm galaxies (6\arcsec\ fields).
Radio centroids are indicated with cross-hairs. 
Redshifts are photometric (submm/radio -- Carilli \& Yun 1999, 2000)
except as indicated with an asterisk
(spectroscopic) or in brackets (the redshift  of the nearby elliptical).
All images except source~7 have been smoothed with a gaussian FWHM 0.05\arcsec\ 
for better visibility.} 
\label{fig1}
\end{figure*}

\begin{figure*}[htb]
\centerline{
\psfig{figure=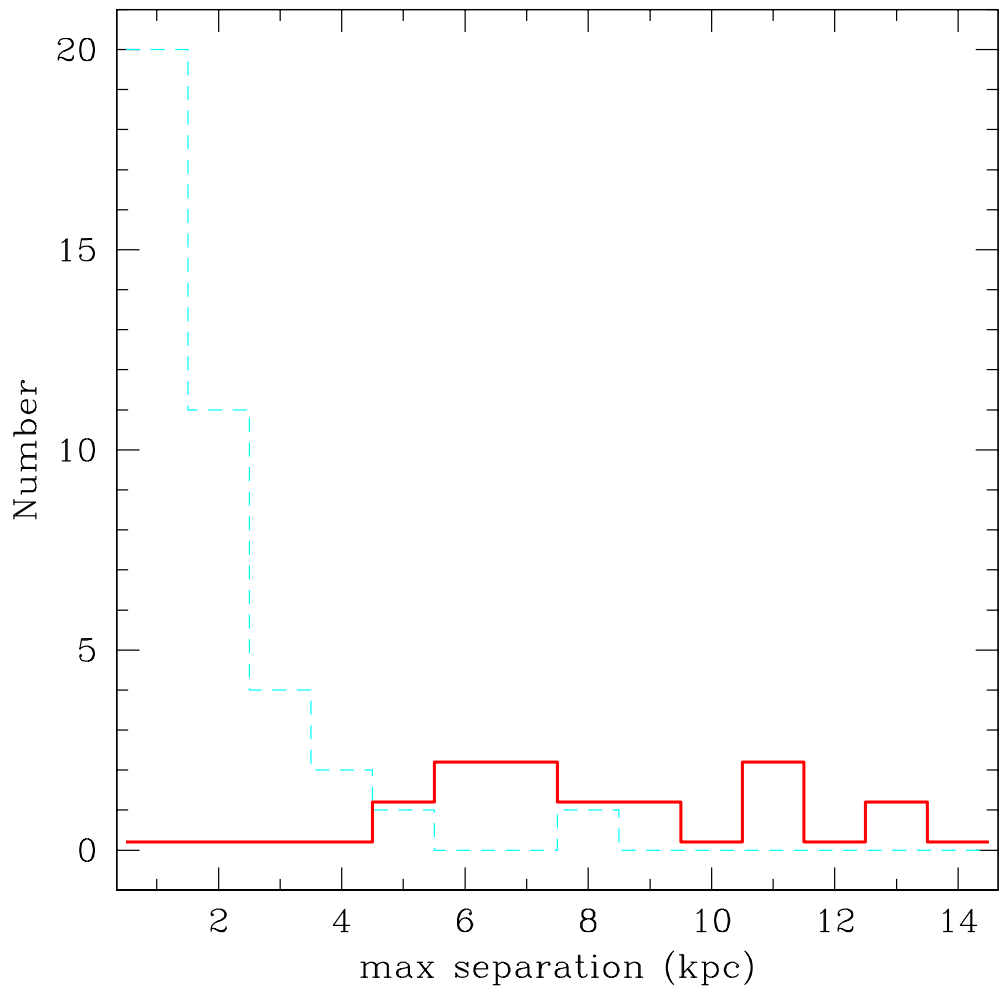,angle=0,width=5.5in}
}
\caption{\footnotesize
Apparent nuclear separations for the submm galaxies (solid line)
compared to the
the 1 Jy sample of {\it IRAS} galaxies from Veilleux, Kim \& Sanders (2002
-- see also Murphy et al.\ 1996) (dashed line).
The submm histogram binning is twice as large as for the IRAS sample.
The distribution of local IRAS galaxies
is highly peaked at small values but has a significant tail at higher values.
The submm galaxies all have separations of $>$5\,kpc except sources (7) and (8)
which do not exhibit multiple component or extended structures.
}
\label{figsep}
\end{figure*}

\begin{figure*}[htb]
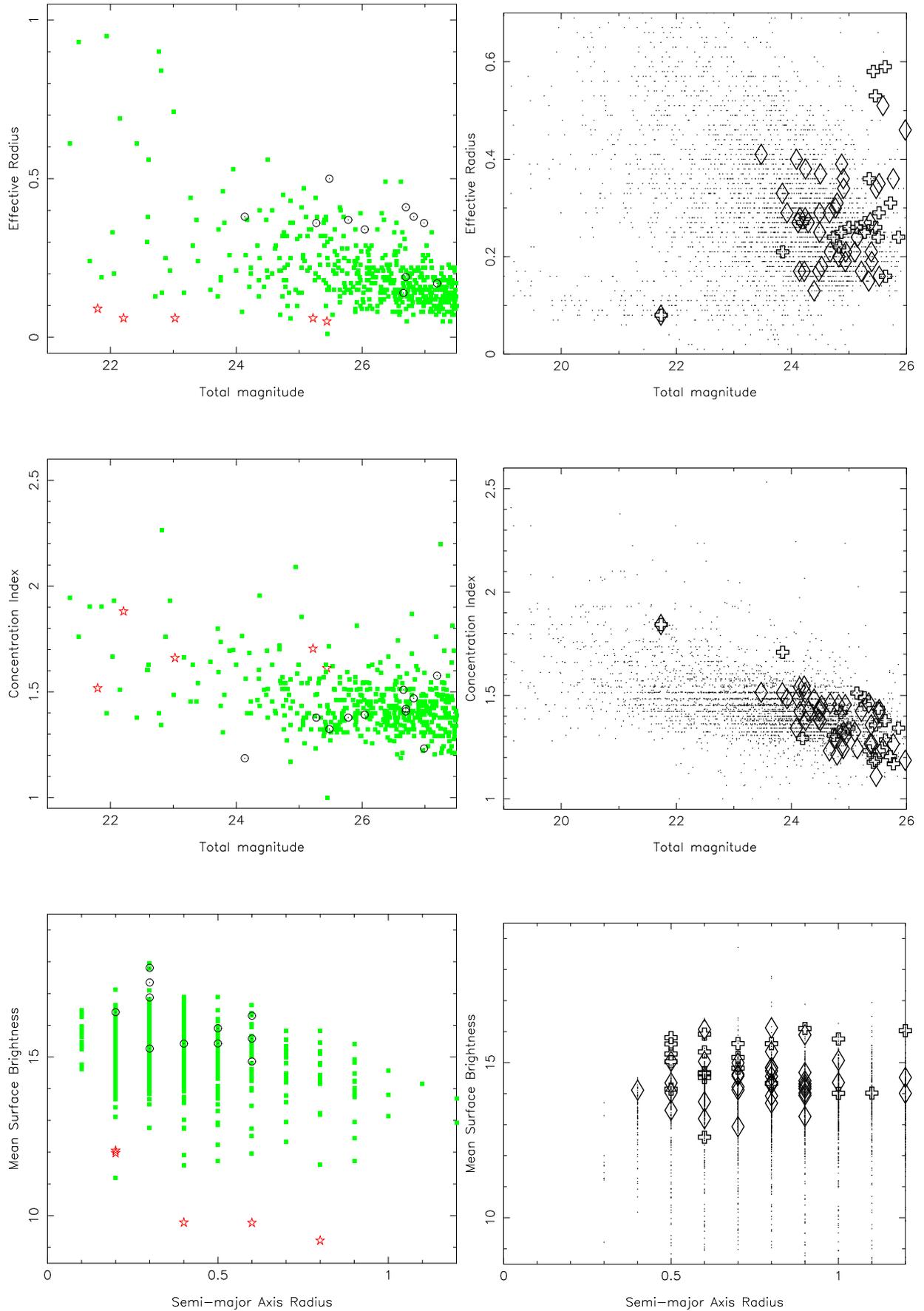

\centerline{
\psfig{figure=f3a.ps,angle=0,width=3.2in}
\psfig{figure=f3d.ps,angle=0,width=3.2in}
}
\vskip1cm
\centerline{
\psfig{figure=f3b.ps,angle=0,width=3.2in}
\psfig{figure=f3e.ps,angle=0,width=3.2in}
}
\vskip1cm
\centerline{
\psfig{figure=f3c.ps,angle=0,width=3.2in}
\psfig{figure=f3f.ps,angle=0,width=3.2in}
}
\caption{\footnotesize
HST-STIS derived morphological parameters for the submm galaxies (left panels)
(submm galaxies -- circles; field galaxies -- squares; stars -- stars).
For comparison, we show the same parameters for the Lyman-break galaxies 
(diamonds) and
$z\sim3.1$ narrow-band Ly$\alpha$ galaxies (crosses) in the right panels.
{\bf Upper row:} effective radius ($R_{50}$) 
	as a function of total magnitude;
{\bf Middle row:} Concentration index ($C32$ = ratio of 75\% to 25\% 
	quartile sizes) vs. magnitude. Higher $C32$ means more compact.
{\bf Lower row:} 
	the mean surface brightness within the effective radius (mag/arcmin$^2$)
	versus semi-major axis radius in arcsec.
}
\label{fig2}
\end{figure*}

\begin{figure*}[htb]
\centerline{
\psfig{figure=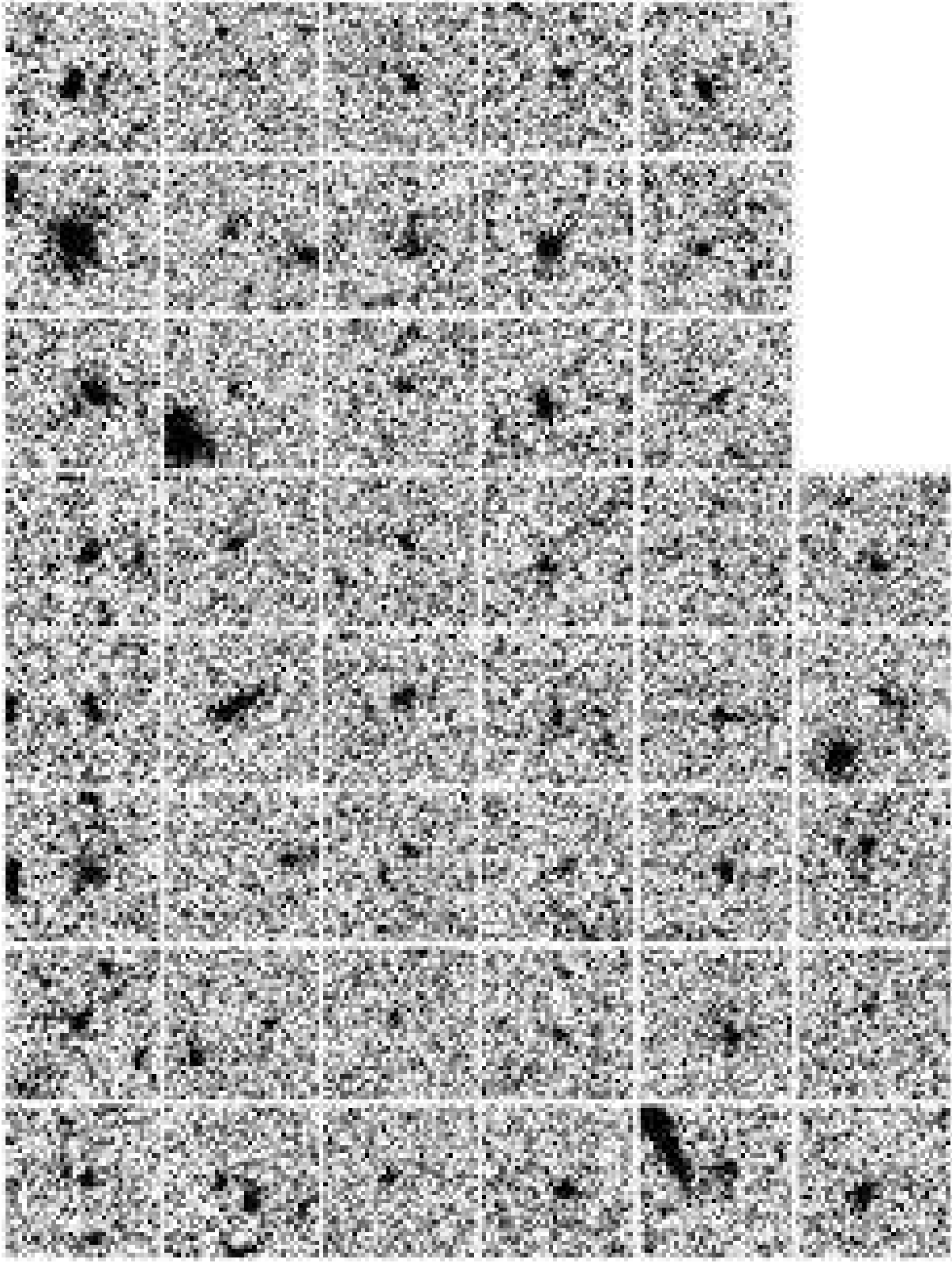,angle=90,width=6.8in}
}
\caption{\footnotesize
{\it HST-WFPC2} observations of Lyman-break galaxies (6\arcsec\ fields).
Galaxies will have redshifts lying within the LBG selection function
of $z$=2.5--3.5 (Steidel et al.\ 1999).
The pixel size is 0.1\arcsec\ for these images. As the sources are generally
brighter than the submm galaxies, and the pixel scale is larger, the images
have been left unsmoothed to preserve extended structure.}
\label{figlbg}
\end{figure*}

\begin{figure*}[htb]
\centerline{
\psfig{figure=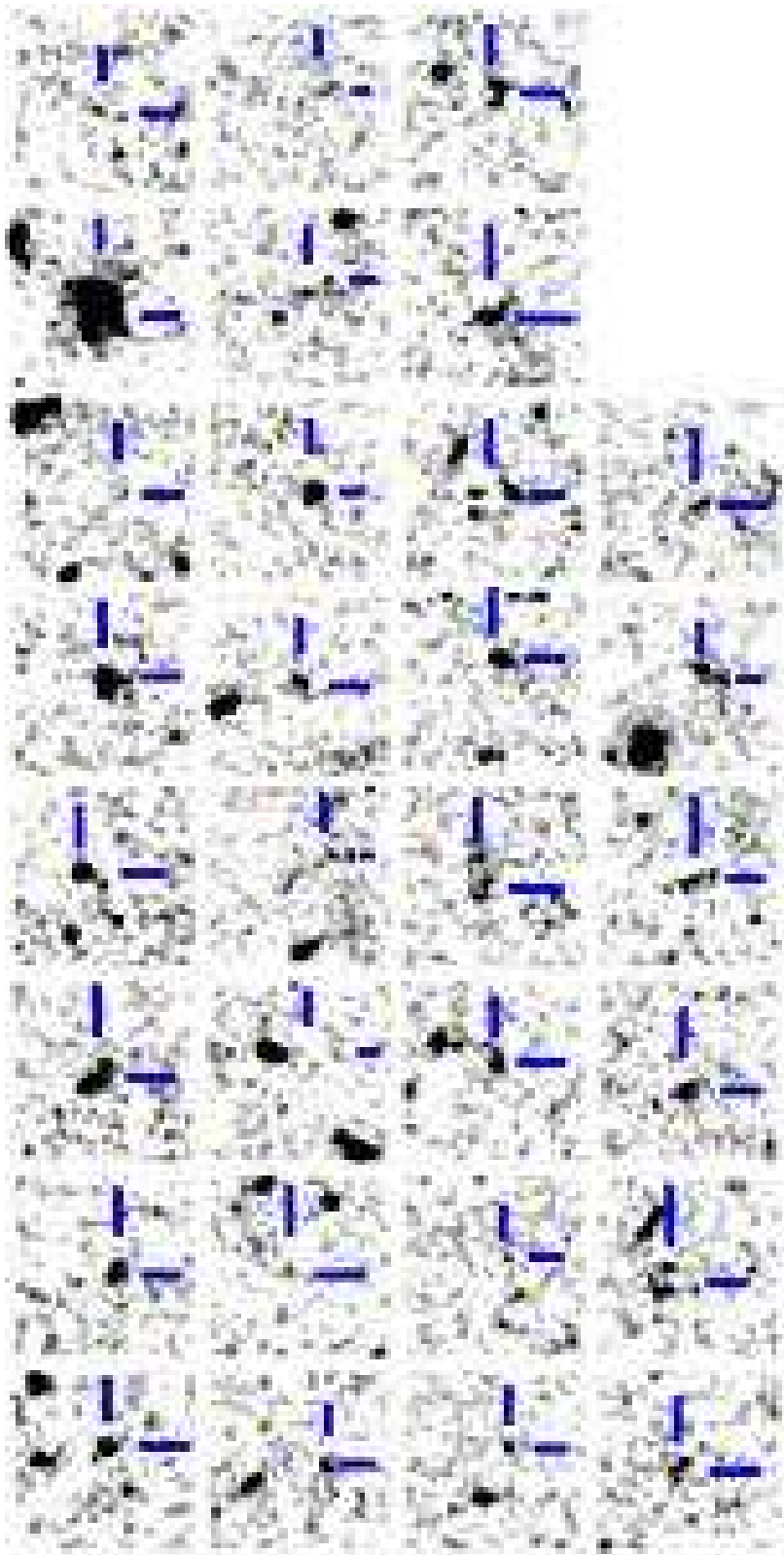,angle=90,width=6.8in}
}
\caption{\footnotesize
{\it HST-WFPC2} observations of $z\sim3.1$ sources selected as narrow-band
Ly$\alpha$ 
excess objects, as described in Steidel et al.\ (2000). Field size is
6\arcsec.
As the narrow-band galaxies  
are generally near the detection limit of the WFPC2 imagery, the 
images have been smoothed with a gaussian FWHM 0.05\arcsec\
for better visibility. The location of the narrow-band source has been indicated with the cross-hatch.}
\label{fignb}
\end{figure*}

\begin{figure*}[htb]
\centerline{
\psfig{figure=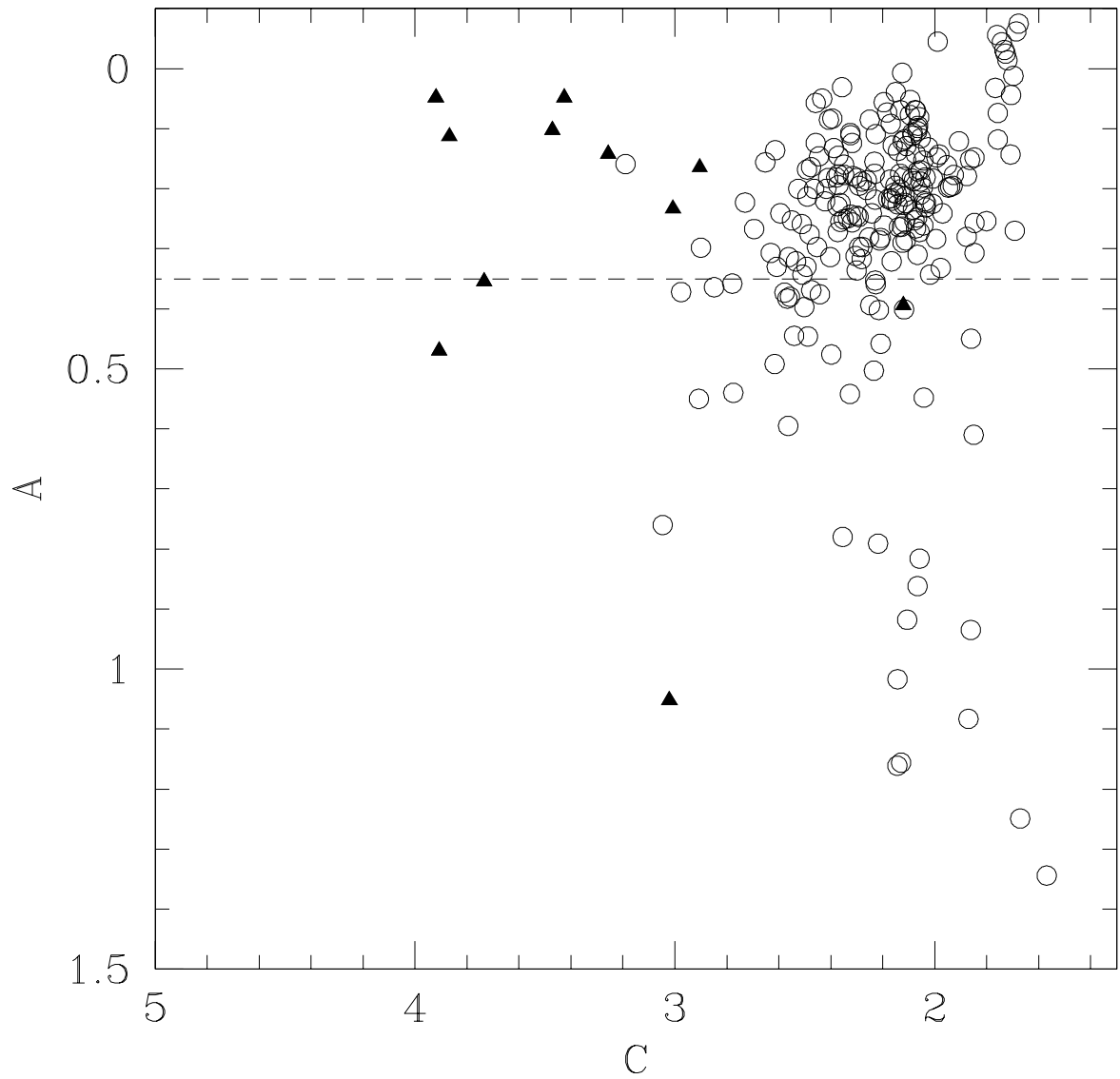,angle=0,width=6.8in}
}
\caption{\footnotesize
A plot of concentration index (C) and asymmetry (A)
for both LBGs in the HDF from $z$=2--3 (circles) and submm galaxies (triangles).
The submm galaxies are significantly more concentrated than the LBGs.
The asymmetries are similar for the submm galaxies 
and LBGs.
}
\label{figac}
\end{figure*}


\begin{deluxetable}{lcccccccccc}
\renewcommand\baselinestretch{1.0}
\tablewidth{0pt} 
\parskip=0.2cm 
\tablenum{1}
\tablecaption{Photometry and Morphology indices for the submm galaxies}
\small
\tablehead{ 
\colhead{source} & S$_{850\mu m}$ & S$_{1.4 GHz}$ &$I$$^a$ &$R'(573)$$^b$ &$R_{50}$$^c$ &$C32$$^d$ &b/a$^e$ &b/a$^f$ &  SB$^g$ &size$^h$ \\
\colhead{} & {(mJy)} & {($\mu$Jy)} &{(mag)} &{(mag)} &{(\arcsec)} &{} &{} &{} &{(mag/\arcsec)} & {(\arcsec)} \\
}
\startdata
(1) SMMJ123553.3+621338 &8.8$\pm$2.1 &58.4$\pm$9.0 &24.84 &25.48 &0.50 &1.32 &0.32 &0.60 &24.80 &0.5  \\ 
(2) SMMJ123600.1+620254 &6.9$\pm$2.0 &262$\pm$17 &24.67 &26.04 &0.34 &1.39 &0.62 &0.34 &25.19 &0.6  \\ 
(3) SMMJ123616.2+621514 &5.8$\pm$1.1 &53.9$\pm$8.4 &24.46 &26.70 &0.41 &1.42 &0.59 &0.71 &26.24 &0.3 \\ 
(4) SMMJ123618.3+621551 &7.8$\pm$1.6 &151$\pm$11 &24.98 &26.66 &0.14 &1.51 &0.88 &0.52 &24.31 &0.4  \\ 
(5) SMMJ123621.3+621708 &7.5$\pm$2.3 &148$\pm$11.0 &23.42 &26.70 &0.19 &1.41 &0.42 &0.53 &24.16 &0.3 \\ 
(6) SMMJ123622.7+621630 &7.1$\pm$1.7 &70.9$\pm$8.7 &24.32 &25.27 &0.36 &1.38 &0.42 &0.58 &24.32 &0.5  \\ 
(7) SMMJ123710.0+622649 &7.4$\pm$2.2 &551$\pm$31 &21.15 &22.43 &0.50 &1.48 &0.63 &0.59 &23.43 &0.7 \\ 
(8) SMMJ123713.9+621827 &15.7$\pm$2.4 &595$\pm$31 &26.80 &27.19 &0.17 &1.58 &1.00 &0.72 &25.30 &0.2 \\ 
(9) SMMJ131231.9+424430 &3.8$\pm$0.8 &127$\pm$7.0 &25.36 &27.00 &0.36 &1.23 &0.86 &0.69 &26.70 &0.3 \\ 
(10) SMMJ131235.2+424424 &3.9$\pm$0.9 &34.1$\pm$7.1 &26.05 &28.03 & & & & & &  \\ 
(11) SMMJ141809.8+522205 &5.4$\pm$1.4 &$<58$ &24.57 &24.14 &0.38 &1.19 &0.70 &0.73 &23.75 &0.6 \\ 
(12) SMMJ221724.7+001242 &13.2$\pm$3.0 &120$\pm$34.0 &25.21 &25.78 &0.37 &1.38 &0.35 &0.75 &24.47 &0.6 \\ 
(13) SMMJ221726.1+001239 &17.8$\pm$2.3 & $<53$ & 26.11 &26.82 &0.38 &1.47 &0.41 &0.27 &25.76 &0.3 \\ 
\enddata
\label{tab1}
(a) Ground based $I$-mag in 3\arcsec\ aperture centered on radio position.\\
\noindent(b) HST mag down to 1.5$\sigma$ ellipse isophot.\\
(c) Effective radius, down to 50\% light contour.\\
(d) Concentration index.\\
(e) $b/a$ from intensity weighted moment.\\ 
(f) $b/a$ from least-squares fit of an ellipse out to the 1.5$\sigma$ SB isophotal contour.\\
(g) mean surface brightness (SB -- mag per arcsec$^2$) within the effective radius.\\ 
(h) image size (in arcsec) from the intensity weighted image moment.
\end{deluxetable}

\begin{deluxetable}{lccc}
\renewcommand\baselinestretch{1.0}
\tablewidth{0pt}
\parskip=0.2cm
\tablenum{1}
\tablecaption{{\cal CAS} morphology indices for the submm galaxies}
\small
\tablehead{
\colhead{source} & concentration & asymmetry & R$_{Petrosian}$ \\
\colhead{} & {} & {} &{(\arcsec)}\\
}
\startdata
(1) SMMJ123553.3+621338 & 3.26 & 0.14 & 2.09 \\ 
(2) SMMJ123600.1+620254 & 3.47 & 0.10 & 1.39 \\ 
(3) SMMJ123616.2+621514 & 3.91 & 0.47 & 1.23 \\ 
(4) SMMJ123618.3+621551 & 3.73 & 0.35 & 1.20 \\ 
(5) SMMJ123621.3+621708 & 3.92 & 0.05 & 4.05 \\ 
(6) SMMJ123622.7+621630 & 3.43 & 0.05 & 2.61 \\ 
(7) SMMJ123710.0+622649 & 3.93 & 0.12 & 0.72 \\ 
(8) SMMJ123713.9+621827 & 3.87 & 0.11 & 0.71 \\ 
(9) SMMJ131231.9+424430 & 2.01 & 0.16 & 1.05 \\ 
(10) SMMJ131235.2+424424 & n/a & n/a & n/a \\ 
(11) SMMJ141809.8+522205 & 3.02 & 1.05 & 0.73 \\ 
(12) SMMJ221724.7+001242 & 3.01 & 0.23 & 7.63 \\ 
(13) SMMJ221726.1+001239 & 2.12 & 0.39 & 0.84 \\ 
\enddata
\label{tab2}
\end{deluxetable}

\begin{deluxetable}{lccc}
\renewcommand\baselinestretch{1.0}
\tablewidth{0pt} 
\parskip=0.2cm 
\tablenum{3}
\tablecaption{Qualitative classifications for submm galaxies, LBGs, and 
narrow-band galaxies}
\small
\tablehead{
\colhead{Classification} & {Submm galaxies} & {LBGs} & {NB-galaxies}\\
}
\startdata
compact regular & 2 (17\%) & 20 (44\%) & 8 (27\%) \\ %
compact irregular & 0 (0\%) & 15 (33\%) & 10 (33\%) \\ %
elongated regular & 0 (8\%) & 2 (30\%) & 1 (3\%) \\ %
elongated irregular & 3 (25\%) & 5 (11\%) & 8 (27\%) \\ %
irregular, multiple component & 7 (58\%) & 3 (7\%) & 3 (10\%) \\ %
\enddata
\label{tab3}
\end{deluxetable}

\end{document}